\PassOptionsToPackage{table}{xcolor}
\documentclass[sigconf]{acmart}

\definecolor{LightGray}{rgb}{0.75,0.75,0.75}

\newcommand{\Ignore}[1]{}

\usepackage{tabularx}
\usepackage{listings}
\usepackage{enumitem}

\AtBeginDocument{%
  \providecommand\BibTeX{{%
    \normalfont B\kern-0.5em{\scshape i\kern-0.25em b}\kern-0.8em\TeX}}}

\setcopyright{acmcopyright}
\copyrightyear{2018}
\acmYear{2018}
\acmDOI{10.1145/1122445.1122456}

\acmConference[ICSE 2022]{The 44th International Conference on Software Engineering}{May 21–29, 2022}{Pittsburgh, PA, USA}

\begin{document}

\title{Software Engineering User Study Recruitment on Prolific:\\An Experience Report}

\author{Brittany Reid, Markus Wagner}
\affiliation{
    \institution{The University of Adelaide}
    \country{Australia}
}
\author{Marcelo d'Amorim}
\affiliation{
    \institution{Universidade Federal de Pernambuco}
    \country{Brazil}
}
\author{Christoph Treude}
\affiliation{
    \institution{The University of Melbourne}
    \country{Australia}
}

\begin{abstract}
Online participant recruitment platforms such as Prolific have been gaining popularity in research, as they enable researchers to easily access large pools of participants.
However, participant quality can be an issue; participants may give incorrect information to gain access to more studies, adding unwanted noise to results.
This paper details our experience recruiting participants from Prolific for a user study requiring programming skills in Node.js, with the aim of helping other researchers conduct similar studies. We explore a method of recruiting programmer participants using prescreening validation, attention checks and a series of programming knowledge questions. 
We received 680 responses, and determined that 55 met the criteria to be invited to our user study. We ultimately conducted user study sessions via video calls with 10 participants. 
We conclude this paper with a series of recommendations for researchers.
\end{abstract}

\settopmatter{printacmref=false}




\maketitle

\section{Introduction}
Prolific is an online recruitment platform with over 150,000 active participants as of December 2021~\cite{prolific}. Members are encouraged to answer demographic questions, which can then be used by researchers to narrow the pool of eligible participants for their studies. However, the existing prescreening features are insufficient; questions are limited and
rely on self-reporting. This self-reporting can be an issue, as Software Engineering user studies need to recruit \emph{programmers}, but non-programmer participants often identify as programmers to be admitted into more studies.

We recruited 10 participants for a user study involving the completion of simple programming tasks in Node.js using a code search tool (NCQ) \cite{reid2021ncq}. While existing work has evaluated the accuracy of self-reported programming skill, establishing its lack of reliability \cite{tahaei2022recruiting}~\cite{doyoucode}, this study documents our experiences recruiting participants for \textbf{user studies} on Prolific. As these types of studies often require researchers to supervise sessions, non-programmer participants can have a significant negative effect, taking up valuable research time.

While a generic prescreening option for programming skills exists, to recruit programmers of a specific language, Prolific recommends~\cite{prolific_custom} first running a custom screening survey to identify the desired demographic, then inviting them to a second study. To recruit Node.js programmers we surveyed 680 participants, using a set of programming knowledge questions derived from Danilova et al.~\cite{doyoucode}. We also used attention checks and validated the answers to prescreening questions and found that 33\% of participants who completed our survey answered inconsistently, providing different answers in our survey than in their prescreening questions.

We identified 55 Node.js programmers from a pool of 206 self-reported Node.js programmers, for a total cost of £193. Our findings suggest
that researchers should not solely rely on self-reported programming skill and instead take measures to verify this information.

\section{Method}

\begin{figure}[!h]
    \centering
    \vspace{-2mm}
    \includegraphics[width=0.97\linewidth]{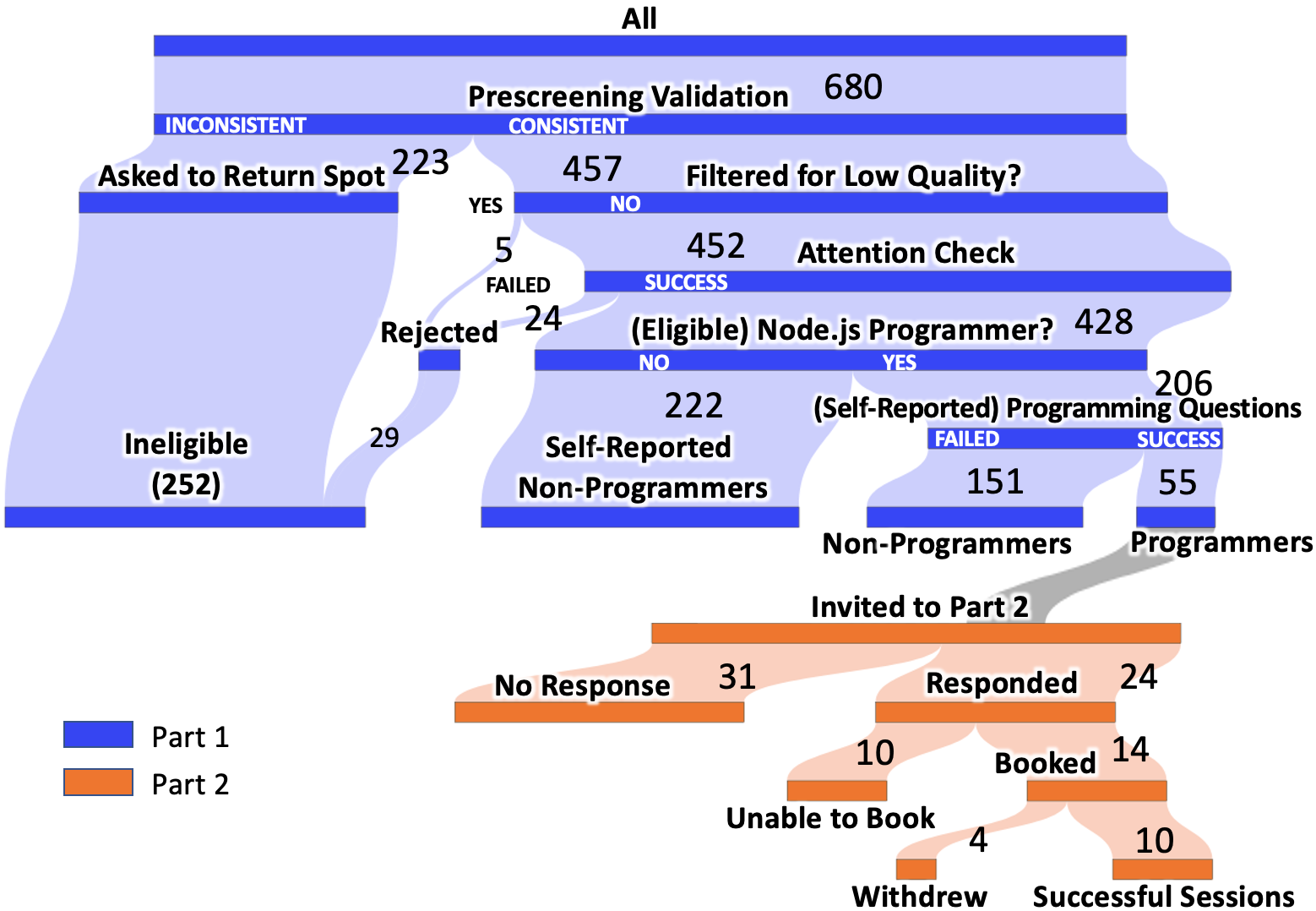}\vspace{-2mm}
    \caption{Diagram of participant recruitment process.}
    \label{fig:process}
\end{figure}

We designed a two-part study consisting of a \emph{screening survey} and \emph{video call} study. All participants who complete Part~1 were paid £0.35, and £7.50 for Part~2. We ran Part~1 in multiple waves with different study sizes, first to pilot the study and then until we reached 10 successful user study sessions. Figure~\ref{fig:process} illustrates the different steps of our participant recruitment process.

All studies on Prolific require a description and link to a survey. Researchers cannot collect any personal details, instead they must collect participant's Prolific IDs. A completion code must be given at the end. We used Google Forms to host our surveys; the full surveys are available on GitHub.\footnote{\url{https://github.com/Brittany-Reid/SE-Prolific-User-Study-Recruitment}}

\subsection{Screening Survey (Part~1 of~2)}
Part~1 was restricted to participants with an approval rating above 95 and who answered yes to the ``programming skills'', ``video call'' and ``English fluency'' prescreeners. In our description, we included the fact this was a two-part study, the pay rate, and a list of requirements including programming skill and ability to execute Node.js code, as participants are not shown what prescreeners are enabled on a study.

The survey had four sections: prescreening validation (P), demographics (D), programming knowledge (K) and the attention check (A). Participants who provided inconsistent prescreening answers were ineligible for our study, and taken to a page of the Google Form asking them to return their place in the study (freeing it for another participant). Prolific requires that all eligible participants in a custom screening survey be paid even if they are not part of the desired demographic, so participants who answered ``Never'' to programming in Node.js (D3) were filtered early, immediately asked to submit and given the completion code. 

\begin{table}[h]
    \centering
    \small
    \caption{Example of programming knowledge questions (K)}\vspace{-3mm}
  \rowcolors{3}{white}{LightGray}
    \begin{tabularx}{\linewidth}{lX}
    \toprule
    \hiderowcolors 
    \showrowcolors
    K5 & Look at the above code, what is the parameter of the function? \\
    K6 & What is the output of the above
code snippet? \\
    K7 & Execute the above code in Node.js,
what is the result? \\
    \bottomrule
    \end{tabularx}
    \label{table:programmingk}
\end{table}

\sloppy
Table~\ref{table:programmingk} shows an example of our programming knowledge questions, adapted 
from Danilova~et~al.~\cite{doyoucode}. We selected the six most effective questions (K1-K6). 
As participants would need to use Node.js in our user study, we also added an additional question (K7), asking participants to demonstrate their ability to execute and provide the output of some Node.js code. We devised a code snippet that could not easily be guessed, but was also easy to read and clearly not malicious (no external libraries or file system access).

To ensure that participants read instructions carefully, an attention check (A1) was also included. Attention check questions ask participants to answer based on some proceeding instructions; in our case, we first told participants to answer the next question with `Java', then asked them `Based on the above text, what is your favourite programming language?'. Attention check failures were rejected.

\subsection{Video Call Booking (Part~2 of~2)}
Participants who answered all programming questions correctly were invited to Part~2 on Prolific. The survey link was used to ask for consent to being recorded, and provide a link to book sessions through YouCanBook.me. Details about the required programming skills and an example programming task were also provided. On booking a session, participants were provided a completion code. We contacted participants who booked a session with our Skype details via Prolific, also reminding them of Prolific's policies on collecting personal information and that they should use their Prolific ID as their username.

\section{Results}


\begin{figure}[!h]
    \centering
    \vspace{-3mm}
    \includegraphics[width=\linewidth]{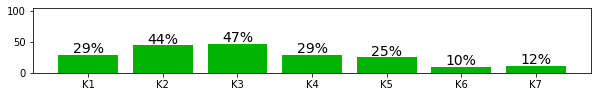}\vspace{-4mm}
    \caption{Percentage of correct answers (non-programmers)}
    \label{fig:answers}
\end{figure}

We received 680 responses to our screening survey, however, there were an additional 664 participants who attempted our study and did not submit before returning their spot or timing out. Participants who returned their spot or timed out did not receive any payment.
Figure~\ref{fig:process} shows the breakdown of participants at each stage of our recruitment process.
We found that 33\% of participants gave inconsistent responses to prescreening validation (P1 and P2), justifying the importance of these questions. Another 24 failed attention checks and 5 were filtered for insufficient effort (nonsensical or random responses), leaving us with 428 eligible participants. 206 participants self-reported as Node.js programmers, and only 55 passed all questions. While only 24 of these participants responded to the invitation to Part~2, this may not reflect total interest; once all places filled, no more participants could respond. 10 participants returned their place without booking, while 14 booked a session. Of these 14, 10 video calls were successful, while 4 withdrew during the session.
Reasons for withdrawal included lack of confidence in programming ability and anxiety programming under observation.

Of the programmer subset, 47\% reported working in a software related industry, compared to just 18\% in the eligible set, indicating that the programming knowledge questions did indeed identify programmers. 85\% of programmers were male, in line with other studies of software engineers \cite{tahaei2022recruiting}. Figure~\ref{fig:answers} shows the percentage of non-programmers who got each question correct; the most effective questions were K6, K7 and K5.

\section{Recommendations}

The results of our screening survey and user study leave us with a set of recommendations for other researchers:

\begin{itemize}[leftmargin=*]
    \item Researchers should not rely on self-reported programming skill when conducting studies. We identified that only 12\% of eligible participants, who self-reported having programming skills using Prolific's prescreening questions, and only 27\% of participants who self-reported as programming in Node.js, met our criteria. We recommend that researchers use other measures such as the programming questions described in our paper.
    \item Prescreener validation is important; 33\% of participants gave inconsistent responses.
    \item Pilot studies are always important; in this case it is especially useful to gather data on the number of non-programmers to account for.
    \item Be clear about the requirements of each study, such as the required programming skill. Participants on Prolific cannot see enabled prescreeners, so we included the need for programming skills in the description for Part 1. In an effort to reduce withdrawal rate, we also included an example task when booking Part 2, so that participants would be confident in meeting our requirements.
\end{itemize}

In summary, finding programmers on Prolific is possible with the right methods, however, the built-in mechanisms are not enough. The method presented in this paper requires a large amount of effort from both researchers and participants, and thus there is still room to improve accurate measures of programming skills on recruitment platforms.


\begin{acks}
This research is supported by an Australian Government Research Training Program (RTP) Scholarship.
\end{acks}

\bibliographystyle{ACM-Reference-Format}
\bibliography{ref}


\begin{thebibliography}{5}


\ifx \showCODEN    \undefined \def \showCODEN     #1{\unskip}     \fi
\ifx \showDOI      \undefined \def \showDOI       #1{#1}\fi
\ifx \showISBNx    \undefined \def \showISBNx     #1{\unskip}     \fi
\ifx \showISBNxiii \undefined \def \showISBNxiii  #1{\unskip}     \fi
\ifx \showISSN     \undefined \def \showISSN      #1{\unskip}     \fi
\ifx \showLCCN     \undefined \def \showLCCN      #1{\unskip}     \fi
\ifx \shownote     \undefined \def \shownote      #1{#1}          \fi
\ifx \showarticletitle \undefined \def \showarticletitle #1{#1}   \fi
\ifx \showURL      \undefined \def \showURL       {\relax}        \fi
\providecommand\bibfield[2]{#2}
\providecommand\bibinfo[2]{#2}
\providecommand\natexlab[1]{#1}
\providecommand\showeprint[2][]{arXiv:#2}

\bibitem[\protect\citeauthoryear{Prolific}{pro}{2021a}]%
        {prolific_custom}
Prolific \bibinfo{year}{2021}\natexlab{a}.
\newblock \bibinfo{booktitle}{\emph{How do I recruit a custom sample?}}
\newblock Prolific.
\newblock
\urldef\tempurl%
\url{https://researcher-help.prolific.co/hc/en-gb/articles/360015732013}
\showURL{%
\tempurl}


\bibitem[\protect\citeauthoryear{Prolific}{pro}{2021b}]%
        {prolific}
Prolific \bibinfo{year}{2021}\natexlab{b}.
\newblock \bibinfo{booktitle}{\emph{Prolific | Online participant recruitment
  for surveys and market research}}.
\newblock Prolific.
\newblock
\urldef\tempurl%
\url{https://www.prolific.co}
\showURL{%
\tempurl}


\bibitem[\protect\citeauthoryear{Danilova, Naiakshina, Horstmann, and
  Smith}{Danilova et~al\mbox{.}}{2021}]%
        {doyoucode}
\bibfield{author}{\bibinfo{person}{Anastasia Danilova}, \bibinfo{person}{Alena
  Naiakshina}, \bibinfo{person}{Stefan Horstmann}, {and}
  \bibinfo{person}{Matthew Smith}.} \bibinfo{year}{2021}\natexlab{}.
\newblock \showarticletitle{Do you Really Code? Designing and Evaluating
  Screening Questions for Online Surveys with Programmers}. In
  \bibinfo{booktitle}{\emph{ICSE}}. \bibinfo{pages}{537--548}.
\newblock


\bibitem[\protect\citeauthoryear{Reid, Barbosa, d'Amorim, Wagner,
  et~al\mbox{.}}{Reid et~al\mbox{.}}{2021}]%
        {reid2021ncq}
\bibfield{author}{\bibinfo{person}{Brittany Reid}, \bibinfo{person}{Keila
  Barbosa}, \bibinfo{person}{Marcelo d'Amorim}, \bibinfo{person}{Markus
  Wagner}, {et~al\mbox{.}}} \bibinfo{year}{2021}\natexlab{}.
\newblock \showarticletitle{NCQ: code reuse support for Node. js developers}.
\newblock \bibinfo{journal}{\emph{arXiv preprint arXiv:2101.00756}}
  (\bibinfo{year}{2021}).
\newblock


\bibitem[\protect\citeauthoryear{Tahaei and Vaniea}{Tahaei and Vaniea}{2022}]%
        {tahaei2022recruiting}
\bibfield{author}{\bibinfo{person}{Mohammad Tahaei} {and} \bibinfo{person}{Kami
  Vaniea}.} \bibinfo{year}{2022}\natexlab{}.
\newblock \showarticletitle{Recruiting Participants With Programming Skills: A
  Comparison of Four Crowdsourcing Platforms and a CS Student Mailing List}. In
  \bibinfo{booktitle}{\emph{CHI}}. \bibinfo{pages}{1–--16}.
\newblock


\end{thebibliography}

\end{document}